# Epitaxially Stabilized Oxide Composed of Twisted Triangular-Lattice Layers

Masaki Uchida,[*,†] Kenta Ohba,[†] Yuki Ohuchi,[†] Yusuke Kozuka,[†] and Masashi Kawasaki[†,‡]

[†]Department of Applied Physics and Quantum-Phase Electronics Center (QPEC), University of Tokyo, Tokyo 113-8656, Japan
[‡]RIKEN Center for Emergent Matter Science (CEMS), Wako 351-0198, Japan

**ABSTRACT:** Layered oxides have been intensively studied due to their high designability for various electronic functions. Here we synthesize a new oxide as epitaxial thin film form by pulsed laser deposition. Film characterizations including cross-section and plan-view transmission electron microscopy confirm that the film is composed of twisted stack of triangular-lattice Rh and Bi layers. We foresee that the concept of twisted oxide layers will open up a new route to design further functional layered oxides.

## INTRODUCTION

Electronic functions and their underlying physics are closely linked to the crystal chemistry. Especially for layered oxides, a wide variety of stacking structures of functional layers have been designed and created in recent decades.[1] One such example is cobalt or rhodium oxides containing CdI$_2$-type CoO$_2$ or RhO$_2$ triangular-lattice layers, attracting attention as thermoelectric materials since the discovery of high thermopower in Na$_x$CoO$_2$.[2-4] A possible origin of the enhanced thermopower is the efficient entropy flow by the correlated electrons in the triangular-lattice layers.[4-6] For exploring novel thermoelectric layered oxides, various compounds have been synthesized, for example by controlling the constituent elements,[7-10] lattice systems,[11-13] lattice strain,[14] and stacking sequence of layers.[15-17]

Here we propose a new degree of freedom in designing layered oxides, i.e. a relative twist of the adjacent layers. Actually, in the research field of atomic layer materials such as graphene or transition metal dichalcogenides, the twist angle has been recognized as a new important parameter for tuning electrical and optical properties of stacked layers.[18-21] In the following, we present a new compound with a twisted stack of oxide layers stabilized by epitaxy technique. For this purpose, we adopt a simple Bi-Rh-O system, because Bi with a relatively large ionic radius is expected to form a large triangular-lattice layer mismatched to the RhO$_2$ layer. A new oxide film composed of the twisted stack of triangular-lattice Rh and Bi layers is grown as a natural superlattice.

## EXPERIMENTAL SECTION

**Film Growth.** Thin films were epitaxially grown by pulsed laser deposition. A ceramic target was prepared by mixing Bi$_2$O$_3$ and Rh$_2$O$_3$ powders at a ratio of 1:1, heating the mixture at 600 ºC for 12 hours in air, and then pelletizing and resintering it at 1000 ºC for 24 hours. For deposition, KrF excimer laser with a wavelength of 248 nm was used to ablate the target. Deposition was performed on (001) SrTiO$_3$ and (111) MgAl$_2$O$_4$ single crystalline substrates, at substrate temperatures of 400-1000 ºC with flowing O$_2$ gas at pressures of $10^{-5}$-3 Torr. The MgAl$_2$O$_4$ substrates were chosen for growing the twisted-layer oxide under the optimum conditions of 800 ºC and 0.3 Torr, while the SrTiO$_3$ ones were used mainly for exploring stable phases over the wide temperature and pressure ranges. Typical laser fluence and repetition rate were 1.4 J/cm$^2$ and 10 Hz, respectively.

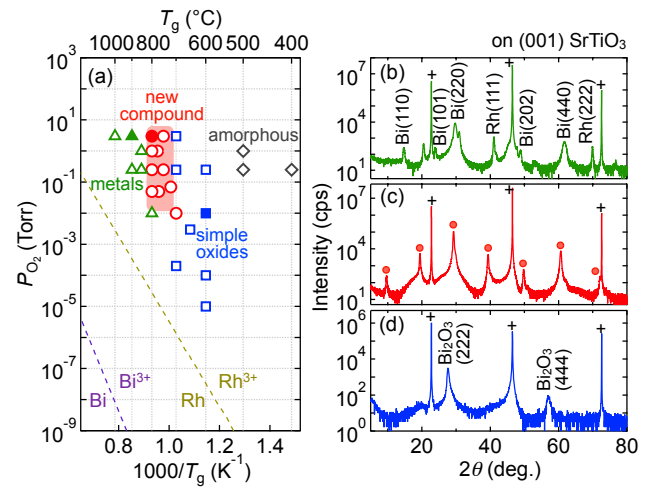

Figure 1. (a) Growth phase diagram of the Bi-Rh-O system as functions of substrate temperature $T_g$ and oxygen pressure $P_{O2}$, classified into four regions based on XRD patterns, as denoted with triangles (metals), circles (a new compound), squares (simple oxides), and diamonds (amorphous). Typical XRD scans for the closed symbols are shown in (b)-(d). Broken lines represent the phase coexistence conditions for the pairs of Bi/Bi$^{3+}$ or Rh/Rh$^{3+}$, taken from the Ellingham diagram.[22] SrTiO$_3$ substrate peaks in the XRD scans are marked with crosses.

**Film Characterization.** After growth, samples were characterized by Rigaku SmartLab x-ray diffractometer, SII Nanocute atomic force microscope, and JEOL JEM-ARM200F scanning transmission electron microscope with energy dis-persive x-ray spectrometer. Longitudinal and Hall resistivities were measured with a four-point probe method in a Quantum Design physical property measurement system. The Seebeck coefficient was measured with a conventional steady-state technique in a cryostat. Optical absorption spectra were obtained by measuring transmission and reflectance at room temperature.

**RESULTS AND DISCUSSION**

Figure 1a summarizes a growth phase diagram of the Bi-Rh-O system, obtained for the films grown on (001) $SrTiO_3$ substrates. The diagram is mainly classified into four regions (metals, a new compound, simple oxides, and amorphous) depending on temperature and oxygen pressure during the film deposition, as exemplified by x-ray diffraction (XRD) patterns shown in Figures 1b-1d. At high temperatures Bi and Rh are reduced into metal, while simple oxides are easily formed at low temperatures. In contrast, only a primary peak and higher harmonics corresponding to a $d$-spacing of 9.15($\pm$0.05) Å, which cannot be found in the ICDD powder diffraction database, appears for the middle narrow growth window of about 700-800 ºC and $10^{-2\text{-}3}$ Torr, as shown in Figure 1c. Considering that this $d$-spacing is close in value to the out-of-plane lattice parameter in trilayer cobalt oxides,[11-12] a similar layered compound is expected to be synthesized in the Bi-Rh-O system. All the bismuth rhodium oxides listed in ICSD database have pyrochlore ($Bi_2Rh_2O_7$), perovskite ($BiRhO_3$), or todorokite ($Bi_6Rh_{12}O_{29}$) structures, and a layered oxide with triangular-lattice composed of Rh and Bi has not been reported yet.

For further optimization and detailed characterization, films were prepared on (111) $MgAl_2O_4$ substrates with triangular-lattice surface. Figure 2a shows out-of-plane XRD scan of the film grown under the optimum conditions. The peaks are assignable to the phase-pure compound with (001) growth orientation. A magnification around the (003) peak in Figure 2b shows clear fringes, and a rocking curve of the film peak in Figure 2c is fairly sharp, demonstrating high-quality epitaxial growth. As shown in Figure 2d, a surface topography taken by atomic force microscopy (AFM) exhibits typical spiral growth, yielding atomically flat step-and-terrace structures with a step height equivalent to the out-of-plane lattice constant. The film with higher crystallinity and surface flatness is obtained on the more lattice-matched triangular-lattice substrate, suggesting that its crystal structure is composed of Rh or Bi triangular-lattice layers.

Analysis of the in-plane atomic arrangement in this new layered compound is not straightforward. In-plane XRD reciprocal space map in Figure 2e reveals a rather complicated six-fold symmetry pattern, which is composed of two triangular-lattice patterns mutually twisted by $2\theta$ ($\theta$=arctan($\sqrt{3}/5$)=19.1º). The in-plane lattice constant of both the patterns is 8.05($\pm$0.05) Å. The result indicates that there are two types of crystallographic domains (A and B) with the in-plane crystal axes misoriented by $\pm\theta$ to the substrate ones. As compared to simulation in Figure 2f, the pattern can be assigned to twisted stack of the Rh and Bi triangular-lattice layers with a relative twist angle of $\pm\theta$, as detailed later. The observed diffraction spots are wholly ascribable to the simulated ones, while their intensity may be dependent on precise atomic positions including oxygen.

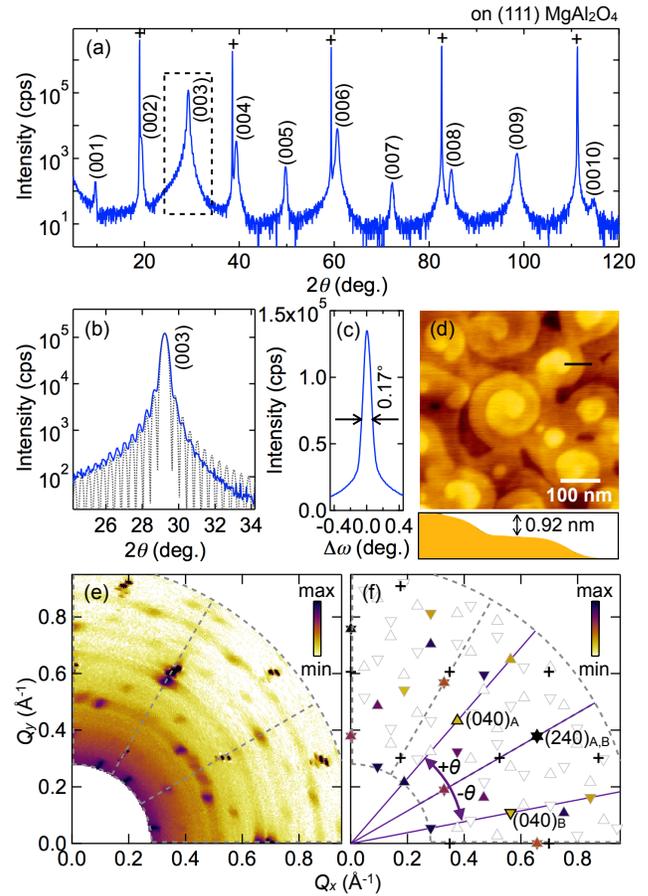

Figure 2. (a) Out-of-plane XRD scan of the new compound film grown on (111) $MgAl_2O_4$ substrate. The substrate peaks are marked with crosses. (b) Magnified view around the (003) film peak. The dotted curve represents a simulation using a film thickness of 23 nm. (c) Rocking curve of the (003) peak with a full width at half maximum of 0.17º. (d) Typical AFM image of the film, with a cross-sectional view along the upper right line. (e) In-plane XRD reciprocal space map, compared to (f) simulated one for twisted stacks of the triangular-lattice Rh and Bi layers with a twist angle of $\pm\theta$=$\pm$19.1º, as illustrated in Figure 3 (k). The upward and downward triangles colored by calculated intensity denote diffraction patterns from two distinct types of crystallographic domains, A and B, respectively. The substrate peak positions are also represented by crosses.

Figure 3a shows a cross-section image of the twisted-layer oxide film, taken by high angle annular dark field (HAADF) scanning transmission electron microscopy (STEM). The STEM image indicates periodic stack of two continuous lines and one dotted line with a period of the out-of-plane $d$-spacing value determined by XRD, as clearly seen in the magnified image in Figure 3b. Energy dispersive x-ray spectrometry (EDX) maps for the same area are given in Figures 3c-3f, clarifying that the continuous lines and the dotted line correspond to Bi bilayer and Rh monolayer, respectively. By carefully examining STEM images, we found a few minor regions with different appearance of the Bi and Rh layers (Figures 3h and 3i), compared with the dominant pattern (Figures 3b and 3g), probably due to the in-plane misarrangement. In the views in Figures 3h and 3i, the Bi bilayer atoms with longer in-plane distances are clearly resolved. The above result indicates that the new compound is composed of the triangular-lattice Rh monolayers and Bi bilayers with different in-plane lattice spacings and orientations.

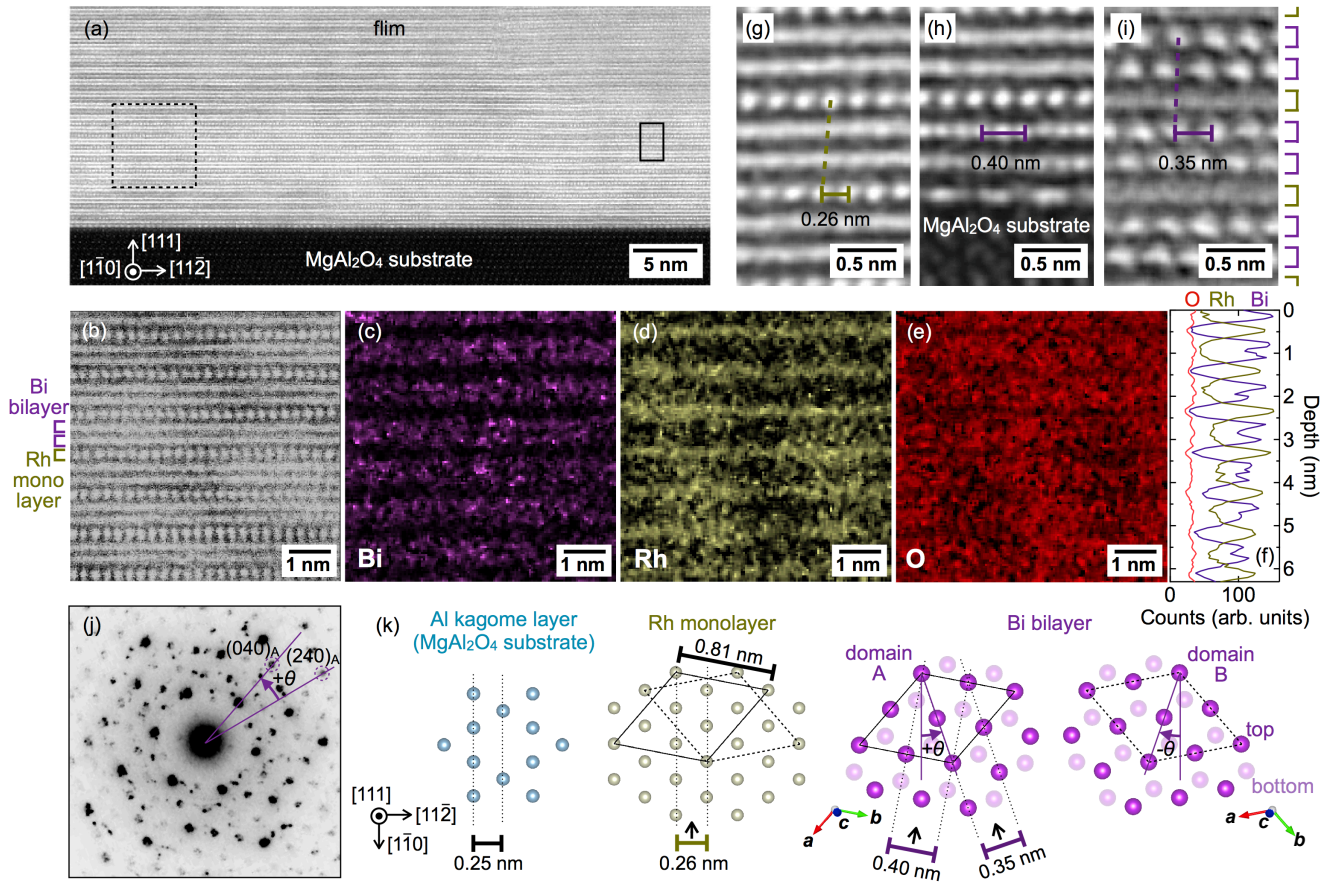

Figure 3. (a) Cross-sectional HAADF-STEM image of the twisted-layer oxide film. (b) Higher-resolution magnified image and corresponding EDX maps for (c) Bi $L$, (d) Rh $L$, and (e) O $K$ edges. (f) Depth profile of Bi, Rh, and O, obtained by integrating the EDX counts along the horizontal direction in (c)-(e). (g)-(i) Other cross-section images demonstrating in-plane atomic arrangement in the respective layers and their stacking pattern, which are the same size as the solid rectangle in (a). (j) Electron diffraction pattern taken in plan-view configuration for a selected area with 150 nm diameter. (k) Top view of the stacking pattern including the MgAl$_2$O$_4$ substrate layer. In the chiral domain structures A and B, the triangular-lattice Rh monolayer and Bi bilayer are alternately stacked with a twist angle of $\pm\theta=\pm19.1°$. The resultant in-plane unit cell is represented by the solid or dotted rhombus. The three arrows correspond to the line of sight for the images in (g)-(i).

In general, two triangular lattices coupled at a twist angle form a quasiperiodic structure that has no unit cell.[18-21] As evidenced by the in-plane x-ray and electron diffractions (Figures 2e and 3j), however, the new layered compound shows periodically modulated hexagonal patterns with the in-plane cell parameter of 8.05($\pm$0.05) Å. The selected-area electron diffraction detects the single pattern from only one type of the two crystallographic domains, while their both patterns are observed in the entire-area in-plane XRD. Actually, owing to the special misorientation angle of $\theta=19.1°$, the triangular-lattice Rh and Bi layers with the different lattice spacings can be commensurately stacked, as illustrated in Figure 3k. While the Rh monolayer is coherently stacked on the lattice-matched substrate layer, the more expanded Bi bilayer is stacked with the commensurate rotation $\pm\theta$. The atomic arrangement is precisely periodic with a unit cell larger than the original Rh and Bi hexagonal unit cells. The domain structures A and B with opposite chirality are formed depending on the sign of the rotation angle, resulting in the observed and simulated in-plane XRD patterns. The projected crystal images shown in Figures 3g-3i are also consistent with this structure model including the length scale.

A schematic crystal structure including possible oxygen coordination, formulated as Bi$_8$Rh$_7$O$_{22}$, is illustrated in Figure 4a. In the twisted-layer oxide, edge-sharing RhO$_6$ octahedra, called the CdI$_2$-type layer, probably form the triangular-lattice Rh monolayer as many other layered rhodium oxides. The determined Rh-Rh distance of about 0.30 nm is also consistent with ones previously reported for the CdI$_2$-type RhO$_2$ layer.[7-10] On the other hand, the Bi bilayer is thought to consist of a (111)-oriented fluorite-type BiO block, considering that Bi-based layered oxides with trigonal symmetry have the similar block structure.[13,23] The stacking pattern in the TEM images in Figure 3 reveals that the $c$-axis is slightly tilted and the twisted-layer oxide has triclinic symmetry with lattice parameters of $a=b=8.05(\pm0.05)$ Å, $c=9.32(\pm0.05)$ Å, $\alpha=83°$, $\beta=87°$, and $\gamma=120°$.

Fundamental electrical and optical properties of the twisted-layer oxide are summarized in Figures 4b-4d. The Seebeck coefficient of 120 $\mu$V/K at room temperature is almost comparable to those of other typical layered rhodium oxides.[7-10,13] On the other hand, the in-plane resistivity shows semiconducting behavior with lowering temperature. Optical absorption spectra give an energy gap of about 1.2 eV, which is close to val-ues in typical $Rh^{3+}$ oxides such as $Rh_2O_3$[24] and $LaRhO_3$.[25] This indicates that the twisted-layer oxide nominally consists of $Bi^{3+}$ and $Rh^{3+}$ oxidation states, as expected from the Ellingham diagram in Figure 1a, and that Rh 4d filled $t_{2g}$ and empty $e_g$ manifolds serve as the valence and conduction bands, respectively. From the hole density of $2\times10^{21}$ cm$^{-3}$ determined by the Hall measurement, the rhodium valence is estimated to be 3.15+, where the total valence in $Bi_8Rh_7O_{22}$ is probably balanced by a slight nonstoichiometry of oxygen or a mixed valence state of bismuth. About 0.15 holes per Rh is consequently doped into the $4d^6$ state, leading to the p-type semiconducting conduction in the $RhO_2$ layers.

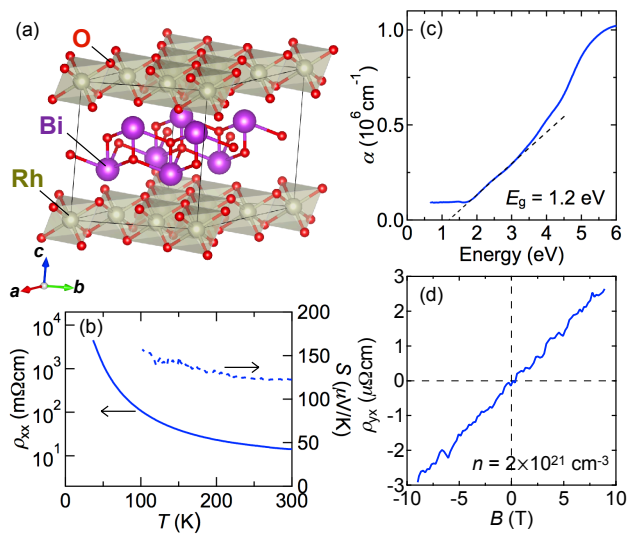

Figure 4. (a) Plausible schematic crystal structure for the twisted-layer oxide. (b) Temperature dependence of the longitudinal in-plane resistivity $\rho_{xx}$ and the thermopower S. (c) Optical absorption spectra with an energy gap of about 1.2 eV, which is derived from linear extrapolation of the absorption edge. (d) Magnetic field dependence of the Hall resistivity $\rho_{yx}$, giving hole density of $2\times10^{21}$ cm$^{-3}$.

Here we have synthesized the new compound with the twisted-layer structure as epitaxial thin film form. Although epitaxy usually refers deposition of crystalline layers with aligned crystal axes, layers in this compound can be epitaxially stacked with a relative twist angle for higher commensurate ordering, because the lattice parameters of the adjacent layers are greatly different and the resulting energy cost needs to be reduced. Our findings open the door to design and create more twisted-layer oxides by controlling the misorientation angle as well as the constituent elements and stacking sequence. For example, rotation angles such as of $\theta$=13.9º (=arctan($\sqrt{3}/7$)), 10.9º (=arctan($\sqrt{3}/9$)) may allow other commensurate stacking of the triangular-lattice layers, with different element ratios and valence states. As a future work, substitution of the constituent elements will also lead to explore new twisted-layer oxides and enhance thermoelectric efficiency through carrier doping.

## CONCLUSIONS

In conclusion, in an effort to propose a new degree of freedom in designing layered oxides, we have grown the high-quality twisted-layer oxide film with using an epitaxy technique. The film characterizations prove that the triangular-lattice Rh monolayers and Bi bilayers with different lattice spacings are alternately stacked with the commensurate twist angle. As other typical thermoelectric rhodium oxides, holes doped into the Rh $4d^6$ state cause p-type conduction and relatively high thermopower. The demonstrated concept of twisted oxide layers will provide a new route to explore further functional layered oxides including thermoelectric materials.

## ASSOCIATED CONTENT

**Supporting Information**

The Supporting Information is available free of charge on the ACS Publications website.

Detailed x-ray diffraction data (PDF).

## AUTHOR INFORMATION

**Corresponding Author**

*M.U. E-mail: uchida@ap.t.u-tokyo.ac.jp

**Notes**

The authors declare no competing financial interest.

## ACKNOWLEDGMENT

The authors gratefully acknowledge insightful discussions with S. Ishiwata, H. Sakai, and A. Tsukazaki. This work was partly supported by Grant-in-Aids for Scientific Research (S) No. 24226002, Young Scientists (A) No. 15H05425, and Challenging Exploratory Research No. 26610098 from MEXT, Japan and by Research Foundation for the Electrotechnology of Chubu. Y. O. acknowledges the support from JSPS through the Program for Leading Graduate Schools (MERIT).

## REFERENCES

(1) Raveau, B. Impact of crystal chemistry upon the physics of strongly correlated electrons in oxides. *Angew. Chem. Int. Ed.* **2013**, 52, 167-175.

(2) Terasaki, I.; Sasago, Y.; Uchinokura, K. Large thermoelectric power in $NaCo_2O_4$ single crystals. *Phys. Rev. B* **1997**, 56, R12685-R12687.

(3) Lee, M.; Viciu, L.; Li, L.; Wang, Y.; Foo, M. L.; Watauchi, S.; Pascal Jr., R. A.; Cava, R. J.; Ong, N. P. Large enhancement of the thermopower in $Na_xCoO_2$ at high Na doping. *Nat. Mater.* **2006**, 5, 537-540.

(4) Wang, Y.; Rogado, N. S.; Cava, R. J.; Ong, N. P. Spin entropy as the likely source of enhanced thermopower in $Na_xCo_2O_4$. *Nature* **2003**, 423, 425-428.

(5) Koshibae, W.; Tsutsui, K.; Maekawa, S. Thermopower in cobalt oxides. *Phys. Rev. B* **2000**, 62, 6869-6872.

(6) Uchida, M.; Oishi, K.; Matsuo, M.; Koshibae, W.; Onose, Y.; Mori, M.; Fujioka, J.; Miyasaka, S.; Maekawa, S.; Tokura, Y. Thermoelectric response in the incoherent transport region near Mott transition: The case study of $La_{1-x}Sr_xVO_3$. *Phys. Rev. B* **2011**, 83, 165127.

(7) Okada, S.; Terasaki, I.; Okabe, H.; Matoba, M. Transport properties and electronic states in the layered thermoelectric rhodate $(Bi_{1-x}Pb_x)_{1.8}Ba_2Rh_{1.9}O_y$. *J. Phys. Soc. Jpn.* **2005**, 74, 1525-1528.

(8) Klein, Y.; Hébert, S.; Pelloquin, D.; Hardy, V.; Maignan, A. Magnetoresistance and magnetothermopower in the rhodium misfit oxide $[Bi_{1.95}Ba_{1.95}Rh_{0.1}O_4][RhO_2]_{1.8}$. *Phys. Rev. B* **2006**, 73, 165121.

(9) Okada, S.; Terasaki, I. Physical properties of Bi-based rhodium oxides with $RhO_2$ hexagonal layers. *Jpn. J. Appl. Phys.* **2005**, 44, 1834-1837.

(10) Shibasaki, S.; Nakano, T.; Terasaki, I.; Yubuta, K.; Kajitani, T. Transport properties of the layered Rh oxide $K_{0.49}RhO_2$. *J. Phys.: Condens. Matter* **2010**, 22, 115603.

(11) Yamauchi, H.; Sakai, K.; Nagai, T.; Matsui, Y.; Karppinen, M. Parent of misfit-layered cobalt oxides: $[Sr_2O_2]_qCoO_2$. *Chem. Mater.* **2006,** 18, 155-158.

(12) Ishiwata, S.; Terasaki, I.; Kusano, Y.; Takano, M. Transport properties of misfit-layered cobalt oxide $[Sr_2O_{2-\nu}]_{0.53}CoO_2$. *J. Phys. Soc. Jpn.* **2006,** 75, 104716.

(13) Kobayashi, W.; Hébert, S.; Pelloquin, D.; Pérez, O.; Maignan, A. Enhanced thermoelectric properties in a layered rhodium oxide with a trigonal symmetry. *Phys. Rev. B* **2007,** 76, 245102.

(14) Brinks, P.; Kuiper, B.; Breckenfeld, E.; Koster, G.; Martin, L. W.; Rijnders, G.; Huijben, M. Enhanced thermoelectric power factor of $Na_xCoO_2$ thin films by structural engineering. *Adv. Energy Mater.* **2014,** 4, 1301927.

(15) Yamamoto, T.; Uchinokura, K.; Tsukada, I. Physical properties of the misfit-layered (Bi,Pb)-Sr-Co-O system: Effect of hole doping into a triangular lattice formed by low-spin Co ions. *Phys. Rev. B* **2002,** 65, 184434.

(16) Hervieu, M.; Boullay, Ph.; Michel, C.; Maignan, A.; Raveau, B. A new family of misfit layered oxides with double rock salt layers $Bi_{\cdot}(A_{0.75\pm}Bi_{0.25\pm}O)_{(3+3x)/2}MO_2$ ($A$=Ca, Sr and $M$=Co, Cr). *J. Solid State Chem.* **1999,** 142, 305-318.

(17) Miyazaki, Y.; Kudo, K.; Akoshima, M.; Ono, Y.; Koike, Y.; Kajitani, T. Low-temperature thermoelectric properties of the composite crystal $[Ca_2CoO_{3.34}]_{0.614}[CoO_2]$. *Jpn. J. Appl. Phys.* **2000,** 39, L531-L533.

(18) Dean, C. R.; Wang, L.; Maher, P.; Forsythe, C.; Ghahari, F.; Gao, Y.; Katoch, J.; Ishigami, M.; Moon, P.; Koshino, M.; Taniguchi, T.; Watanabe, K.; Shepard, K. L.; Hone, J.; Kim, P. Hofstadter's butterfly and the fractal quantum Hall effect in moiré superlattices. *Nature* **2013,** 497, 598-602.

(19) Ponomarenko, L. A.; Gorbachev, R. V.; Yu, G. L.; Elias, D. C.; Jalil, R.; Patel, A. A.; Mishchenko, A.; Mayorov, A. S.; Woods, C. R.; Wallbank, J. R.; Mucha-Kruczynski, M.; Piot, B. A.; Potemski, M.; Grigorieva, I. V.; Novoselov, K. S.; Guinea, F.; Fal'ko, V. I.; Geim, A. K. Cloning of Dirac fermions in graphene superlattices. *Nature* **2013,** 497, 594-597.

(20) Hunt, B.; Sanchez-Yamagishi, J. D.; Young, A. F.; Yankowitz, M.; LeRoy, B. J.; Watanabe, K.; Taniguchi, T.; Moon, P.; Koshino, M.; Jarillo-Herrero, P.; Ashoori, R. C. Massive Dirac fermions and Hofstadter butterfly in a van der Waals heterostructure. *Science* **2013,** 340, 1427-1430.

(21) Heo, H.; Sung, J. H.; Cha, S.; Jang, B.-G.; Kim, J.-Y.; Jin, G.; Lee, D.; Ahn, J.-H.; Lee, M.-J.; Shim, J. H.; Choi, H.; Jo, M.-H. Interlayer orientation-dependent light absorption and emission in monolayer semiconductor stacks. *Nat. Commun.* **2014,** 6, 7372.

(22) Ellingham, H. J. T. Transactions and Communications. *J. Soc. Chem. Ind.* **1944,** 63, 125-160.

(23) Conflant, P.; Boivin, J. C.; Thomas, D. J. Etude structurale du conducteur anionique $Bi_{0.765}Sr_{0.235}O_{1.383}$. *J. Solid State Chem.* **1980,** 35, 192-199.

(24) Koffyberg, F. P. Optical bandgaps and electron affinities of semiconducting $Rh_2O_3(I)$ and $Rh_2O_3(III)$. *J. Phys. Chem. Solids* **1992,** 53, 1285-1288.

(25) Nakamura, M.; Krockenberger, Y.; Fujioka, J.; Kawasaki, M.; Tokura, Y. Perovskite $LaRhO_3$ as a $p$-type active layer in oxide photovoltaics. *Appl. Phys. Lett.* **2015,** 106, 072103.